\documentclass[9pt,conference]{IEEEtran}
\usepackage[preprint]{dcase2026}

\usepackage{bm} %

\usepackage[numbers,sort&compress]{natbib}

\usepackage{dcase2026,amsmath,graphicx,url,times,booktabs, tabularx, pgfplots, hyperref, tikz, float, xcolor}
\pgfplotsset{compat=1.18}
\usetikzlibrary{positioning,arrows.meta,fit,backgrounds,calc}

\definecolor{cPre}{RGB}{120,120,120}
\definecolor{cFeat}{RGB}{125,95,185}   %
\definecolor{cClass}{RGB}{40,150,120}  %
\definecolor{cPost}{RGB}{225,140,40}   %

\definecolor{cBMABZ}{RGB}{0,114,178}    %
\definecolor{cD}{RGB}{213,94,0}         %
\definecolor{cBP}{RGB}{204,121,167}     %
\definecolor{cF1}{RGB}{64,64,64}        %

\title{NAVE: A Normalized, Adaptive Conformer for \\ Antarctic Whale Vocalization Detection}

\name{Matthias Nagl$^{1}$,
      Paul Primus$^{1}$,
      Gerhard Widmer$^{1,2}$}
\address{$^1$Institute of Computational Perception, Johannes Kepler University Linz, Austria \\
$^2$LIT Artificial Intelligence Lab, Linz, Austria\\
matthias.nagl@jku.at\\
}

\begin{document}

\maketitle

\begin{abstract}
Antarctic whale vocalization detection requires temporally localizing call boundaries in long underwater recordings and classifying them into predefined call types. Training reliable machine learning models for this task is difficult due to the variability and sparsity of whale calls in passive acoustic monitoring datasets.
These difficulties are further amplified by time-varying noise that can mask whale calls and by site-specific ambient sound characteristics to which a trained model may not generalize.
In this work, we present NAVE, a Conformer-based detector that improves whale-call detection via three components. First, its input representation combines per-bin complex mean subtraction, which suppresses constant site-specific ambient sound characteristics, with an additional per-channel energy normalization (PCEN) channel that adapts to temporal variations in the noise floor. Second, a frequency-dynamic stem processes these noise-aware representations with convolutions that adapt across the characteristic frequency bands of the target call types. Third, we use a Conformer backbone with a convolution kernel matched to the duration of the target calls that adds an inductive bias and improves temporal localization.
We evaluate NAVE on BioDCASE 2026 Task 2 using leave-one-site-out
cross-validation across the validation sites, which approximates deployment on
unseen recording sites. A single NAVE model with a tuned per-class threshold exceeds the stronger WhaleVAD-BPN baseline, using no boundary-proposal network and no search over post-processing parameters. A controlled ablation study shows that the normalized front-end and a duration-matched convolutional kernel each contribute substantially to recovering the D downsweep, the hardest class for previous detectors, with the wide kernel giving the largest aggregate gain.
\end{abstract}


\section{Introduction}
\label{sec:intro}

Passive acoustic monitoring (PAM) provides a non-invasive means of studying Antarctic blue and fin whales, whose wide-ranging movements across remote oceans make direct observation challenging \cite{Cooke2018Blue, Cooke2018Fin}. It produces large volumes of low-signal-to-noise-ratio (SNR) audio whose manual annotation is slow and requires expertise \cite{Kowarski2021Review}. Automatic detection is therefore valuable but difficult: calls are sparse, conditions vary across sites, and call appearance changes with distance, noise, and instrumentation.

\begin{figure}[t]
\centering
\includegraphics[width=\columnwidth]{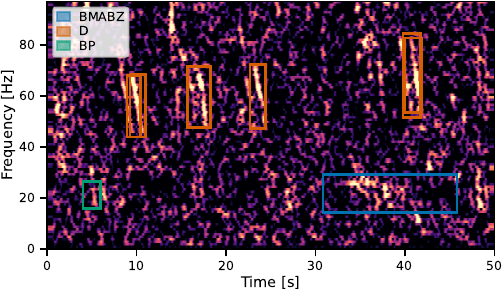}
\caption{Spectrogram with whale calls annotated by time--frequency boxes. Each box marks an event's extent and call type: BMABZ, D, or BP.}
\label{fig}
\end{figure}

BioDCASE 2026 Task 2 targets automatic detection of Antarctic blue- and fin-whale calls in long PAM recordings. Models must generalize across sites and soundscapes while detecting sparse events, making it a useful benchmark for whale-call sound event detection systems \cite{JeanLabadye2025BioDCASE, Miller2021Dataset}. The task uses strong multi-label annotations: systems predict each call's category and temporal bounds, as shown in Figure~\ref{fig}. Seven call types are grouped by acoustic similarity into three evaluation classes~\cite{Schall2024Benchmark}: BMABZ for blue-whale A-, B-, and Z-calls, D for blue- and fin-whale downsweeps, and BP for fin-whale $20$\,Hz pulses. The development set comes from the Acoustic Trends Blue Fin Library (ATBFL)~\cite{Miller2020ATBFL, Miller2021Dataset}.

Two dataset properties shaped our design. First, calls occupy little of the audio, causing severe temporal imbalance. Even modest false-positive rates can therefore yield many spurious detections, making false-positive control as important as recall. Second, class prevalence varies strongly across site-years. D-calls are especially uneven, with most training examples from \texttt{elephantisland2013} and \texttt{elephantisland2014}, and are the rarest class at the validation sites. D-calls have also remained the main bottleneck for prior detectors, with the lowest per-class F1 under both Whale-VAD and WhaleVAD-BPN \cite{Geldenhuys2025WhaleVAD, Geldenhuys2025WhaleVADBPN}.

Existing detectors such as Whale-VAD~\cite{Geldenhuys2025WhaleVAD} commonly use a CNN-BiLSTM backbone and improve it through output gating or tuned post-processing, as in WhaleVAD-BPN \cite{Geldenhuys2025WhaleVADBPN}. We ask whether the classification backbone and front-end offer similar opportunities for performance improvements, and present NAVE: a Conformer~\cite{Gulati2020Conformer} with a normalized, adaptive front-end and a depthwise convolution matched to call duration. A single NAVE model exceeds WhaleVAD-BPN. Ablations show that the normalized front-end and duration-matched kernel each recover much of the performance on D-calls, although it remains the hardest class. Our contributions are:
\begin{itemize}
\item A noise-aware representation combining per-bin complex mean subtraction~\cite{Geldenhuys2025WhaleVAD} with a fixed per-channel energy normalization (PCEN) channel~\cite{Lostanlen2019PCEN, Wang2017PCEN}, substantially improving faint D-call detection.
\item A frequency-dynamic stem \cite{Nam2022FDY} whose convolutions adapt across class-specific frequency bands instead of sharing one kernel across all frequencies.
\item A depthwise Conformer kernel matched to call duration, with an analysis of when further widening stops helping \cite{Luo2016ReceptiveField}.
\end{itemize}

Code is available at \url{https://github.com/TheFruitGuy/NAVE-SED}.

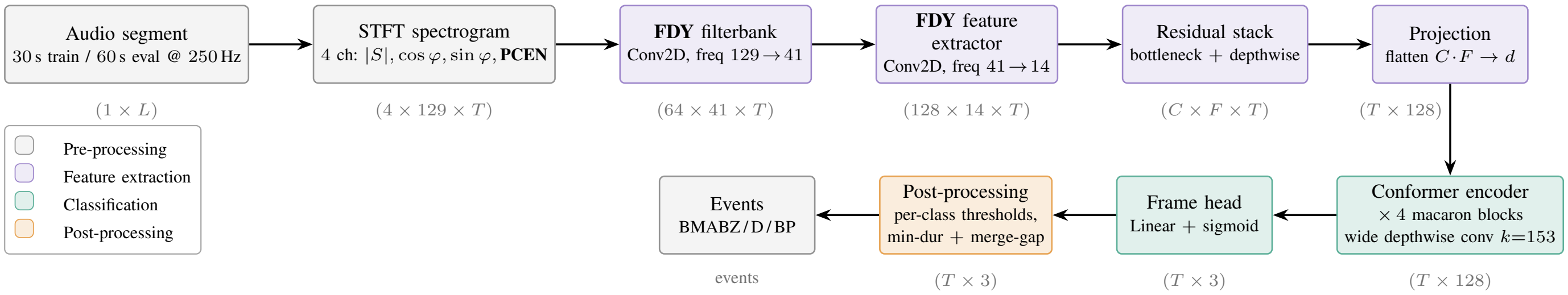
\begin{figure*}[t]
\centering
\resizebox{\textwidth}{!}{%
\begin{tikzpicture}[
  font=\small,
  >={Stealth[length=2.2mm]},
  mod/.style={draw, rounded corners=2pt, align=center, minimum height=11mm,
              minimum width=22mm, inner sep=3pt, line width=0.6pt,
              font=\footnotesize},
  pre/.style={mod, draw=cPre!80, fill=cPre!8},
  feat/.style={mod, draw=cFeat!75, fill=cFeat!12},
  class/.style={mod, draw=cClass!75, fill=cClass!14},
  post/.style={mod, draw=cPost!80, fill=cPost!16},
  flow/.style={->, line width=0.8pt},
  dimlbl/.style={font=\scriptsize, text=black!55},
]

\node[pre]  (audio) {Audio segment\\{\scriptsize $30$\,s train / $60$\,s eval @ $250$\,Hz}};
\node[pre,  right=9mm of audio] (stft)
  {STFT spectrogram\\{\scriptsize $4$ ch: $|S|,\cos\varphi,\sin\varphi,$\,\textbf{PCEN}}};
\node[feat, right=9mm of stft] (fb)
  {\textbf{FDY} filterbank\\{\scriptsize Conv2D, freq $129\!\to\!41$}};
\node[feat, right=9mm of fb] (fe)
  {\textbf{FDY} feature\\extractor\\{\scriptsize Conv2D, freq $41\!\to\!14$}};
\node[feat, right=9mm of fe] (rs)
  {Residual stack\\{\scriptsize bottleneck $+$ depthwise}};
\node[feat, right=9mm of rs] (proj)
  {Projection\\{\scriptsize flatten $C\!\cdot\!F\to d$}};

\foreach \a/\b in {audio/stft, stft/fb, fb/fe, fe/rs, rs/proj}
  \draw[flow] (\a) -- (\b);

\node[dimlbl, below=1.5mm of audio] {$(1\times L)$};
\node[dimlbl, below=1.5mm of stft]  {$(4\times 129\times T)$};
\node[dimlbl, below=1.5mm of fb]    {$(64\times 41\times T)$};
\node[dimlbl, below=1.5mm of fe]    {$(128\times 14\times T)$};
\node[dimlbl, below=1.5mm of rs]    {$(C\times F\times T)$};
\node[dimlbl, below=1.5mm of proj, xshift=-7mm] {$(T\times 128)$};

\node[class, below=13mm of proj] (enc)
  {Conformer encoder\\{\scriptsize $\times\,4$ macaron blocks}\\{\scriptsize wide depthwise conv $k{=}153$}};
\node[class, left=9mm of enc] (head)
  {Frame head\\{\scriptsize Linear $+$ sigmoid}};
\node[post,  left=9mm of head] (pp)
  {Post-processing\\{\scriptsize per-class thresholds,}\\{\scriptsize min-dur $+$ merge-gap}};
\node[pre,   left=9mm of pp] (events)
  {Events\\{\scriptsize BMABZ\,/\,D\,/\,BP}};

\draw[flow] (proj.south) -- (enc.north);          %
\foreach \a/\b in {enc/head, head/pp, pp/events}
  \draw[flow] (\a) -- (\b);

\node[dimlbl, below=1.5mm of enc]    {$(T\times 128)$};
\node[dimlbl, below=1.5mm of head]   {$(T\times 3)$};
\node[dimlbl, below=1.5mm of pp]     {$(T\times 3)$};
\node[dimlbl, below=1.5mm of events] {events};

\node[draw=black!35, rounded corners=2pt, line width=0.5pt, anchor=south west,
      inner sep=4pt, font=\scriptsize] at (audio.west |- events.south) {%
  \begin{tabular}{@{}cl@{}}
    \tikz\filldraw[fill=cPre!8,  draw=cPre!80]  (0,0) rectangle (2.6mm,2.6mm); & Pre-processing\\[1pt]
    \tikz\filldraw[fill=cFeat!12, draw=cFeat!75] (0,0) rectangle (2.6mm,2.6mm); & Feature extraction\\[1pt]
    \tikz\filldraw[fill=cClass!14,draw=cClass!75](0,0) rectangle (2.6mm,2.6mm); & Classification\\[1pt]
    \tikz\filldraw[fill=cPost!16, draw=cPost!80] (0,0) rectangle (2.6mm,2.6mm); & Post-processing\\
  \end{tabular}%
};

\end{tikzpicture}%
}
\vspace{-6mm}
\caption{Overview of NAVE. A four-channel phase-aware spectrogram (including a
fixed PCEN channel) feeds an FDY convolutional stem and a four-block Conformer
with a wide depthwise kernel ($k{=}153$). A per-class sigmoid head yields frame
posteriors that are post-processed into events.}
\label{fig:system}
\end{figure*}

\section{Related Work}
\label{sec:related}

Among the systems developed for the whale call detection task in previous editions of the BioDCASE, Whale-VAD~\cite{Geldenhuys2025WhaleVAD} and its extension WhaleVAD-BPN~\cite{Geldenhuys2025WhaleVADBPN} provide strong task-specific baselines. Whale-VAD, the earlier of the two systems, introduced a CNN-BiLSTM backbone comprising a convolutional feature extractor, a recurrent sequence model, and a per-class frame-level prediction head. It also uses a trigonometric phase-aware spectrogram representation that improves detection of low-frequency baleen-whale calls. WhaleVAD-BPN, our stronger baseline, retains the same acoustic backbone and adds a boundary-proposal network that gates the frame-level posteriors, together with a systematic search over post-processing parameters. Its improvement over Whale-VAD therefore comes from gating the classifier output and tuning post-processing, rather than from the acoustic model itself.

The components used in NAVE have been established in other domains. The Conformer~\cite{Gulati2020Conformer} combines self-attention with convolution to model global and local temporal structure jointly. PCEN~\cite{Lostanlen2019PCEN, Wang2017PCEN} replaces log compression with adaptive gain control, reducing the influence of slowly varying background noise. Frequency-dynamic convolution (FDY)~\cite{Nam2022FDY} relaxes the assumption of a single shared kernel across frequency, allowing filters to adapt to different frequency regions. These properties are well suited to faint, band-specific calls embedded in non-stationary noise. NAVE integrates these components for whale-call detection, with a particular focus on the front-end and acoustic backbone.

\section{Method}
\label{sec:method}
This section describes the main components of NAVE: the normalized front-end, the frequency-dynamic stem, the Conformer backbone, training, and post-processing. Figure~\ref{fig:system} provides an overview of the complete system. %

\subsection{Normalized, adaptive front-end}
\label{ssec:frontend}

Each recording is divided into fixed-length segments of $30$\,s. For each segment, we compute a short-time Fourier transform (STFT) using a $256$-sample Hann window (approximately $1$\,s at the $250$\,Hz sampling rate) and a $5$-sample hop ($20$\,ms). This yields a complex spectrogram with $129$ frequency bins. The long analysis window provides the resolution needed for the low-frequency vocalizations of baleen whales.

From the complex spectrogram, we construct a three-channel phase-aware representation. For each frequency bin, the complex-valued mean over time is subtracted within each segment~\cite{Geldenhuys2025WhaleVAD}. This removes persistent spectral components while retaining temporally varying call structure. The resulting complex spectrogram is represented by its magnitude and by the cosine and sine of its phase. This trigonometric phase representation follows Whale-VAD~\cite{Geldenhuys2025WhaleVAD}, where it was shown to improve the detection of low-frequency whale calls.

We append a fourth channel obtained by applying PCEN to the raw magnitude spectrogram~\cite{Lostanlen2019PCEN, Wang2017PCEN}. PCEN combines adaptive gain control with compressive nonlinear scaling. For each frequency bin, PCEN estimates the local background energy using an exponential moving average with smoothing coefficient $s=0.025$. The input magnitude is then divided by this estimate with gain-control strength $\alpha=0.98$, followed by dynamic-range compression with bias $\delta=2.0$ and exponent $r=0.5$. This frequency-wise adaptation reduces the influence of slowly varying background energy and increases the contrast of transient events relative to the local noise floor.
The PCEN channel is intended particularly to improve sensitivity to faint D-call downsweeps that are partially masked by background noise. It enhances such events before feature extraction, although it may also amplify acoustically similar transients. The resulting four-channel representation is passed to the convolutional stem.

\subsection{Frequency-dynamic stem}
\label{ssec:fdy}

NAVE's convolutional stem converts the four-channel spectrogram into a per-frame feature sequence, following the overall structure of the Whale-VAD feature extractor~\cite{Geldenhuys2025WhaleVAD}. Figure~\ref{fig:system} shows the stem as a sequence of purple boxes. A filterbank convolution first maps the $129$ input frequency bins to a smaller set of learned frequency bands. A two-layer feature-extraction block then further compresses the frequency axis while increasing the channel dimension, after which a bottleneck convolution reduces the number of channels. Finally, a short stack of depthwise convolutions aggregates information over local time--frequency neighborhoods. The resulting feature map is flattened across the channel and frequency dimensions and projected to the Conformer model width. Across the stem, the frequency axis is reduced from $129$ bins to $14$, with most of this reduction occurring in the early layers, where detailed frequency information is still available. The time dimension is not downsampled, so both the Conformer and the frame-level head operate at a $20$\,ms frame rate.

The filterbank convolution and the first convolution of the feature-extraction block use FDY convolution~\cite{Nam2022FDY}. The remaining layers use standard two-dimensional convolutions. FDY is motivated by the fact that absolute frequency is class-informative for these whale calls, so the same local pattern may require different filters in different frequency bands. FDY represents the effective kernel at each output frequency as a mixture of $K=4$ learned basis kernels, with mixture weights obtained by softmax attention over the temporally averaged input. We apply FDY only in the earliest layers, where the frequency resolution is still high and adaptation to the characteristic bands of the call classes is most useful.

\subsection{Conformer backbone and its receptive field}
\label{ssec:backbone}

NAVE uses a four-block Conformer encoder~\cite{Gulati2020Conformer} as its temporal backbone. Each block contains two feed-forward modules, one before and one after the self-attention and convolution sublayers, following the macaron configuration. The encoder has a model width of $128$ and uses four attention heads with rotary positional embeddings~\cite{Su2024RoFormer}. A frame-level linear prediction head maps the encoder output to three class logits at the original temporal resolution of one frame every $20$\,ms. The complete model contains $2{,}705{,}787$ parameters.

The convolution module in each Conformer block contains a depthwise temporal convolution whose kernel width determines the amount of contiguous local context processed by a single layer. At the $20$\,ms frame rate, the adopted kernel width of $153$ frames corresponds to $3.06$\,s. Since self-attention already provides global context across the input segment, the purpose of the wide convolution is not to extend the model's temporal reach. Instead, it introduces a fixed, local, translation-equivariant operation over a window long enough to span a whale call, complementing the global and content-dependent interactions learned by self-attention.

Because the convolution is depthwise, increasing its width adds only $d_{\mathrm{model}} \times k$ parameters per block, or approximately $19{,}600$ weights for $k=153$. The local receptive field can therefore be enlarged with little effect on the total model size. We treat the kernel width as the main architectural variable of the backbone and evaluate it systematically in Section~\ref{ssec:dev}. Odd kernel widths are used so that symmetric padding preserves the number of frames.

\subsection{Training}
\label{ssec:training}

We optimize the model with RAdam~\cite{Liu2020RAdam}, which is designed to reduce the need for learning-rate warmup. The complete training configuration is given in Table~\ref{tab:config}. Following Whale-VAD~\cite{Geldenhuys2025WhaleVAD}, we use a masked, weighted binary cross-entropy loss: padded frames are excluded, and the positive term for each class $c$ is scaled by $w_c = N/N_c$, where $N_c$ is the number of positive segments for class $c$ and $N$ is the total number of positive segments across all three classes. This gives greater weight to the rarer classes. Because calls are sparse, each epoch pairs all positive segments with an equal number of randomly sampled no-call segments, with the negative set resampled at the start of every epoch.

We maintain an exponential moving average (EMA) of the model weights and batch-normalization statistics~\cite{Izmailov2018SWA}, and use the averaged weights for validation and checkpointing. This stabilizes training against the per-epoch resampling of the negative pool, which can otherwise shift the model's score distributions and the resulting precision--recall trade-off between epochs, particularly for the sparse D-call class.  Because we rely on weight averaging, we use a constant learning rate rather than a decaying schedule, as averaging generally benefits from a non-decaying learning rate~\cite{Izmailov2018SWA}. At the end of each epoch, the per-class detection thresholds are re-tuned as described in Section~\ref{ssec:postproc}, and the checkpoint is selected according to the resulting F1 score.

\subsection{Post-processing}
\label{ssec:postproc}
At inference, each recording is divided into overlapping $60$\,s segments with a $2$\,s overlap. Frame-level posteriors are averaged in the overlapping regions to produce a single posterior sequence for each class over the full recording. We then convert these sequences into events using the fixed post-processing pipeline of Whale-VAD~\cite{Geldenhuys2025WhaleVAD}. The longer $60$\,s inference segments, compared with the $30$\,s segments used during training, provide additional context for each prediction.

Each posterior sequence is smoothed with a $0.5$\,s median filter and thresholded using a class-specific detection threshold. Consecutive events of the same class are merged when separated by less than $0.5$\,s, and events shorter than $0.5$\,s or longer than $30$\,s are discarded. Only the per-class detection thresholds are tuned, using coordinate descent. The smoothing window, merge gap, and duration limits are fixed and shared across all classes.

\section{Experimental Setup}
\label{sec:setup}

We use the BioDCASE Task 2 development set, drawn from the ATBFL introduced in Section~\ref{sec:intro}. Eight site-years form the training set: \texttt{ballenyisland2015}, \texttt{casey2014}, \texttt{elephantisland2013}, \texttt{elephantisland2014}, \texttt{greenwich2015}, \texttt{kerguelen2005}, \texttt{maudrise2014}, and \texttt{rosssea2014}. Three site-years form the validation set used for checkpoint selection and threshold tuning: \texttt{casey2017}, \texttt{kerguelen2014}, and \texttt{kerguelen2015}. The training configuration is given in Table~\ref{tab:config}.

\begin{table}[h]
\centering
\caption{Training configuration.}
\label{tab:config}
\begin{tabularx}{\columnwidth}{@{}lX@{}}
\toprule
Hyperparameter & Value \\
\midrule
Optimizer            & RAdam \\
Learning rate        & $5\times10^{-5}$, constant, no warmup \\
Weight decay         & $10^{-3}$ \\
Gradient clipping    & $1.0$ \\
Batch size           & 32 \\
Training epochs      & 40 \\
Loss                 & Masked weighted BCE, $w_c=N/N_c$ \\
Segment length       & $30$\,s \\
Negative sampling    & 1:1 with positive segments; resampled each epoch \\
Weight averaging     & EMA, decay $0.999$ \\
Dropout              & $0.1$ \\
Checkpoint selection & Validation F1 after threshold tuning \\
\bottomrule
\end{tabularx}
\end{table}

Scores are computed at the event level. Predicted events are matched to the reference annotations using one-dimensional temporal intersection over union (IoU) with a threshold of $0.3$. For each of the $C=3$ classes, the resulting true-positive, false-positive, and false-negative counts define precision $P_c$ and recall $R_c$. We use the primary challenge metric, macro F1, which first averages precision and recall across classes and then computes their harmonic mean:
\begin{equation}
  \label{eqn:f1}
  \mathrm{F1} = \frac{2\bar{P}\bar{R}}{\bar{P}+\bar{R}}, \qquad
  \bar{P} = \frac{1}{C}\sum_{c=1}^{C} P_c, \qquad
  \bar{R} = \frac{1}{C}\sum_{c=1}^{C} R_c.
\end{equation}

All results are evaluated using a leave-one-site-out (LOSO) threshold-selection protocol over the three validation site-years. The trained model remains fixed across folds. For each held-out site, the class-specific detection thresholds are selected on the other two sites and then applied to the held-out site. The event counts from the three held-out evaluations are pooled before computing
$P_c$ and $R_c$ and the overall F1 in Eq.~\ref{eqn:f1}, as in the official evaluation. The baselines instead average precision and recall across folds~\cite{Geldenhuys2025WhaleVADBPN}. Each configuration is trained with five random seeds, and we report the mean and sample standard deviation across seeds.

\section{Results}
\label{sec:results}

We evaluate NAVE in two stages. Section~\ref{ssec:dev} assesses the contribution of each design choice through a controlled ablation, with particular attention to the recovery of D-calls through the normalized front-end and duration-matched kernel. Section~\ref{ssec:loso} then compares the resulting single-model system with the published Whale-VAD baselines.

\subsection{Development}

\label{ssec:dev}

Table~\ref{tab:main} reports the mean per-class F1 across five seeds, together with the mean and sample standard deviation of the overall macro F1. The standard deviation of macro F1 ranges from approximately $0.01$ to $0.02$. We therefore interpret small differences cautiously and base our conclusions primarily on the larger improvements and their class-wise patterns. The overall ranking of the main design choices is stable across seeds.

\begin{table}[t]
  \centering
  \caption{Leave-one-site-out comparison and ablation on the three validation
site-years using macro F1 (Eq.~\ref{eqn:f1}). The upper block contains the
published Whale-VAD baselines~\cite{Geldenhuys2025WhaleVAD,
Geldenhuys2025WhaleVADBPN}; the lower block adds one NAVE component per row, all rows sharing the mean-subtracted Whale-VAD input representation. Five-seed means, with the sample standard deviation of F1 in
parentheses.}
  \label{tab:main}
  \footnotesize
  \setlength{\tabcolsep}{5pt}
  \begin{tabular}{@{}lcccc@{}}
    \toprule
    System & BMABZ & D & BP & F1 \\
    \midrule
    Whale-VAD~\cite{Geldenhuys2025WhaleVAD, Geldenhuys2025WhaleVADBPN} & {0.628} & 0.126 & 0.315 & 0.377 \\
    WhaleVAD-BPN~\cite{Geldenhuys2025WhaleVADBPN} & 0.615 & 0.340 & 0.460 & 0.475 \\
    \midrule
    Conformer base                   & 0.393 & 0.100 & 0.465 & 0.328 (0.019) \\
    \quad $+$ FDY stem               & 0.472 & 0.155 & 0.479 & 0.376 (0.009) \\
    \quad $+$ PCEN channel           & 0.400 & 0.269 & 0.507 & 0.399 (0.016) \\
    \quad $+$ wide kernel ($k{=}65$) & 0.466 & 0.383 & 0.516 & 0.463 (0.009) \\
    NAVE ($k{=}153$)                 & 0.547 & {0.438} & {0.518} & {0.518} (0.016) \\
    \quad fold-averaged              & 0.581 & 0.449 & 0.463 & 0.510 (0.012) \\
    \bottomrule
  \end{tabular}
\end{table}

The second section of Table~\ref{tab:main} adds one component at a time to the Conformer architecture. Adding the FDY stem raises macro F1 from $0.328$ to $0.376$, driven mainly by BMABZ, with a smaller gain for D-calls and little change for BP. Adding the PCEN channel produces the first substantial gain for D-calls, raising their F1 from $0.155$ to $0.269$. This improvement comes with a decrease in BMABZ from $0.472$ to $0.400$, but the gain on D-calls and BP increases the overall F1 from $0.376$ to $0.399$. PCEN's gain control tracks the background over roughly $0.8$\,s ($s=0.025$ at a $20$\,ms hop), so it normalizes sustained BMABZ tonals toward the floor while shorter D and BP calls outrun it.

Introducing the wide depthwise kernel gives the largest single improvement in the ablation. At $k=65$, all three classes improve and macro F1 rises from $0.399$ to $0.463$. Increasing the kernel width further to the adopted value of $k=153$ raises macro F1 to $0.518$, with further gains for BMABZ and D-calls.

\begin{figure}[t]
  \centering
  \begin{tikzpicture}
    \begin{axis}[
      width=\columnwidth, height=0.68\columnwidth,
      xlabel={Depthwise kernel size $k$}, ylabel={LOSO F1},
      xmin=28, xmax=164, ymin=0.22, ymax=0.60,
      xtick={31,49,65,81,97,113,129,145,161},
      ytick={0.25,0.30,0.35,0.40,0.45,0.50,0.55,0.60},
      tick label style={font=\scriptsize},
      label style={font=\footnotesize},
      legend style={font=\scriptsize, at={(0.5,1.02)}, anchor=south,
                    legend columns=4, draw=none, /tikz/every even column/.append style={column sep=4pt}},
      legend cell align=left,
      grid=major, grid style={gray!18, line width=0.2pt},
      every axis plot/.append style={line width=0.9pt, mark size=1.1pt,
        /pgfplots/error bars/y dir=both,
        /pgfplots/error bars/y explicit,
        /pgfplots/error bars/error mark=none,
        /pgfplots/error bars/error bar style={line width=0.4pt}},
      clip=false,
    ]
      \draw[gray!75, dashed, line width=0.8pt] (axis cs:153,0.22) -- (axis cs:153,0.60);
      \node[font=\tiny, anchor=south east, gray!95] at (axis cs:153,0.225) {$k{=}153$};

      \addplot[cBMABZ, mark=*] coordinates {
        (31,0.438) +- (0,0.051) (49,0.485) +- (0,0.028) (57,0.455) +- (0,0.035)
        (65,0.466) +- (0,0.025) (73,0.510) +- (0,0.007) (81,0.466) +- (0,0.048)
        (89,0.509) +- (0,0.049) (97,0.497) +- (0,0.044) (105,0.500) +- (0,0.037)
        (113,0.511) +- (0,0.045) (121,0.520) +- (0,0.035) (129,0.523) +- (0,0.043)
        (137,0.559) +- (0,0.024) (145,0.558) +- (0,0.035) (153,0.547) +- (0,0.029)
        (161,0.561) +- (0,0.012)};
      \addlegendentry{BMABZ}
      \addplot[cD, mark=triangle*] coordinates {
        (31,0.260) +- (0,0.022) (49,0.297) +- (0,0.015) (57,0.353) +- (0,0.043)
        (65,0.383) +- (0,0.031) (73,0.404) +- (0,0.040) (81,0.423) +- (0,0.027)
        (89,0.380) +- (0,0.028) (97,0.433) +- (0,0.018) (105,0.454) +- (0,0.022)
        (113,0.424) +- (0,0.030) (121,0.422) +- (0,0.042) (129,0.400) +- (0,0.033)
        (137,0.414) +- (0,0.042) (145,0.434) +- (0,0.017) (153,0.438) +- (0,0.024)
        (161,0.423) +- (0,0.042)};
      \addlegendentry{D}
      \addplot[cBP, mark=square*] coordinates {
        (31,0.505) +- (0,0.018) (49,0.515) +- (0,0.015) (57,0.515) +- (0,0.005)
        (65,0.516) +- (0,0.013) (73,0.522) +- (0,0.009) (81,0.523) +- (0,0.015)
        (89,0.522) +- (0,0.017) (97,0.522) +- (0,0.009) (105,0.527) +- (0,0.014)
        (113,0.519) +- (0,0.003) (121,0.517) +- (0,0.007) (129,0.522) +- (0,0.008)
        (137,0.514) +- (0,0.004) (145,0.515) +- (0,0.012) (153,0.518) +- (0,0.013)
        (161,0.499) +- (0,0.012)};
      \addlegendentry{BP}
      \addplot[cF1, mark=diamond*, line width=1.5pt,
               /pgfplots/error bars/error bar style={line width=0.5pt}] coordinates {
        (31,0.409) +- (0,0.011) (49,0.436) +- (0,0.002) (57,0.446) +- (0,0.007)
        (65,0.463) +- (0,0.009) (73,0.486) +- (0,0.014) (81,0.481) +- (0,0.020)
        (89,0.482) +- (0,0.020) (97,0.494) +- (0,0.024) (105,0.504) +- (0,0.021)
        (113,0.497) +- (0,0.027) (121,0.498) +- (0,0.015) (129,0.492) +- (0,0.010)
        (137,0.511) +- (0,0.014) (145,0.517) +- (0,0.015) (153,0.518) +- (0,0.016)
        (161,0.508) +- (0,0.014)};
      \addlegendentry{macro}
    \end{axis}
  \end{tikzpicture}
  \caption{Effect of the depthwise kernel width on per-class and macro F1 for the full PCEN$+$FDY architecture using $60$\,s inference segments. Points show the mean, error bars one standard deviation, over at least three seeds. Macro F1 rises sharply up to approximately $k=65$ and then enters a broad upper plateau, from which the adopted $k=153$ (dashed) is taken.}
  \label{fig:kernel}
\end{figure}
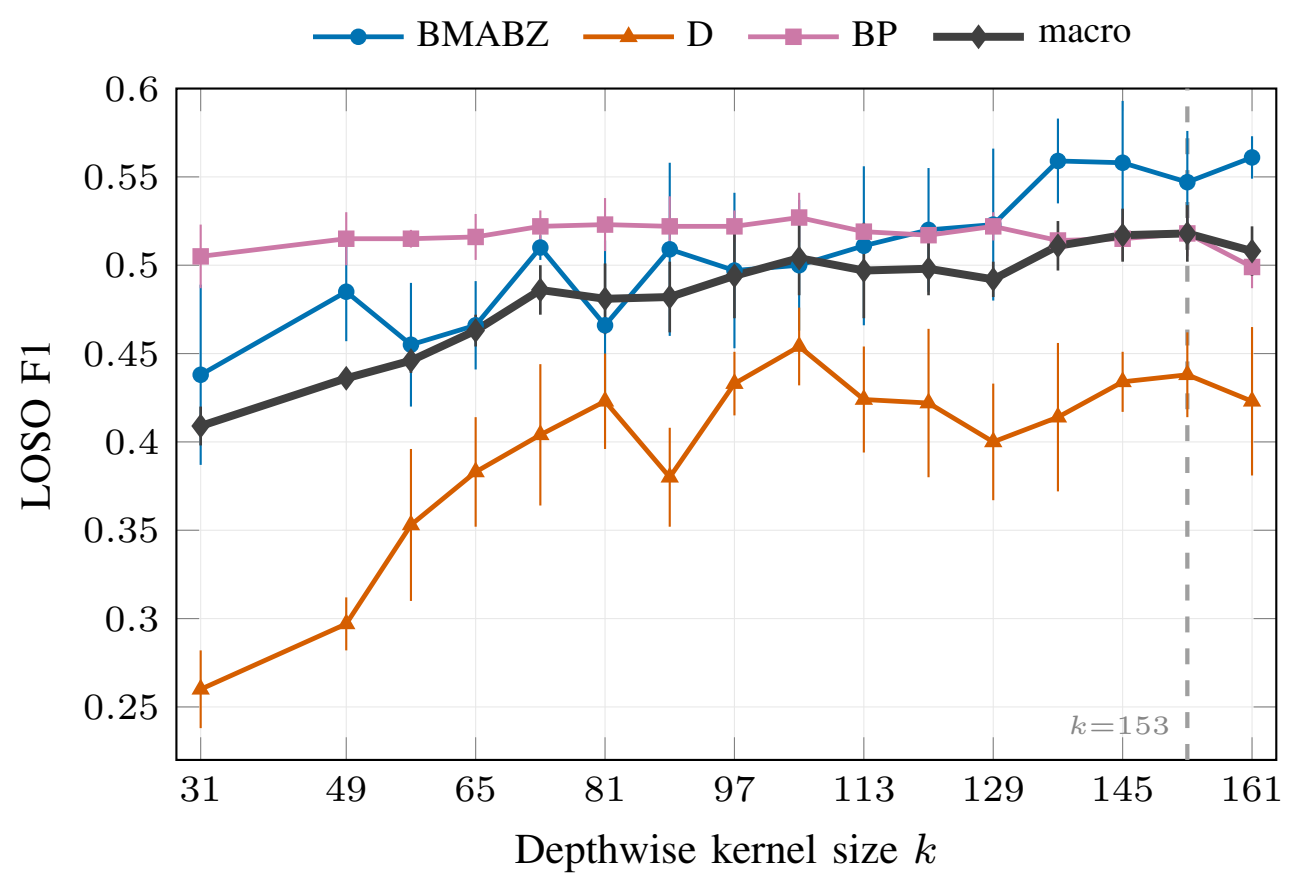

Figure~\ref{fig:kernel} examines the effect of kernel width in more detail while holding the remainder of the architecture and training procedure fixed. Macro F1 rises considerably up to approximately $k=65$, corresponding to a temporal span of about $1.3$\,s, and then varies across a broad upper plateau. The improvement beyond $k=65$ is driven primarily by BMABZ, whose performance generally increases with kernel width. BP is comparatively insensitive to widening, while D-call performance improves over the narrowest kernels but fluctuates within the seed variation across the upper range.

These class-wise trends are consistent with a relationship between receptive field size and call duration: widening helps until the kernel spans a call, so BMABZ, which contains the longest calls, keeps gaining, whereas the shorter BP calls already perform strongly with narrow kernels. We adopt $k=153$, corresponding to $3.06$\,s of local temporal context. It improves macro F1 by $0.026$ over $k=129$, better on all five paired seeds (paired $t$-test, $p=0.036$). Several kernels on the upper plateau lie within the seed variation, so $k=153$ is not necessarily optimal, but it is a well-supported choice from the best-performing region.

We attribute the improved detection of D-calls to both the front-end and the backbone: in Table~\ref{tab:main} the PCEN channel and the wide kernel each contribute a large, comparable gain. D-calls nevertheless remain the weakest of the three classes, and the kernel sweep shows diminishing returns once the local receptive field is sufficiently wide. This suggests that further gains are more likely to come from richer or learned input representations and additional labeled sites than from increased backbone capacity.

\subsection{Generalization and comparison}
\label{ssec:loso}

Table~\ref{tab:main} compares NAVE with the two Whale-VAD baselines under the
leave-one-site-out protocol of Section~\ref{sec:setup}. Both baselines are
reported in \cite{Geldenhuys2025WhaleVADBPN} under three-fold cross-validation
over the same validation site-years and with the same macro F1 metric, though
that work averages precision and recall across folds where we pool event counts. A single NAVE model reaches a pooled macro F1 of 0.518 (0.016) over five
seeds, above both baselines, including WhaleVAD-BPN at $0.475$. All five seeds
exceed this baseline, with the lowest reaching $0.502$. Using the baselines'
aggregation, which averages precision and recall across the three folds rather
than pooling counts, NAVE reaches $0.510$, so it leads under either convention.

NAVE reaches this with a single forward pass and class-specific thresholds,
without the boundary-proposal stage or post-processing search of WhaleVAD-BPN.
The gain is driven by D-calls, raised from $0.340$ to $0.449$, while BP is
essentially unchanged ($0.463$ against $0.460$) and BMABZ falls modestly
($0.581$ against $0.615$). A redesigned front-end and acoustic backbone can
therefore outperform a stronger pipeline built around the fixed CNN-BiLSTM model.

\section{Conclusion}
\label{sec:conclusion}

We presented NAVE, a single Conformer-based detector for Antarctic blue- and fin-whale calls with a normalized, adaptive front-end. Under leave-one-site-out cross-validation, NAVE outperforms both Whale-VAD and WhaleVAD-BPN using a single forward pass and class-specific detection thresholds, without a boundary-proposal stage or post-processing search. Its advantage is driven primarily by improved detection of D-calls, which benefit from both the front-end and the duration-matched kernel, and outweighs a modest reduction on BMABZ.

Two directions follow from these results. First, D-call performance improves substantially with both the normalized front-end and the duration-matched kernel, but remains the weakest once the receptive field is sufficiently wide. Since the preferred kernel widths also broadly reflect call duration, a natural next step is to introduce class-specific modeling throughout the system, for example through a multi-resolution front-end combined with class-specific receptive fields. Second, bridging the gap between held-out cross-validation and deployment at genuinely unseen sites will likely require explicit cross-site adaptation, such as stronger front-end normalization, test-time adaptation, and additional labeled data from a broader range of sites.

\section{Acknowledgment}
\label{sec:ack}
We thank Florian Schmid and Tobias Morocutti for helpful discussions, and
Christiaan Geldenhuys for his correspondence and for releasing Whale-VAD, an
inspiration for the Conformer developed here. The LIT AI Lab is supported by the
Federal State of Upper Austria.

\bibliographystyle{IEEEtran}
\bibliography{references}

\end{document}